\documentclass[a4paper, 11pt, notitlepage]{article}
\usepackage[utf8]{inputenc}
\usepackage{epsfig}
\usepackage{amssymb}
\usepackage{amsmath}
\usepackage{float}

\usepackage{hyperref}

\begin{document}

\title{Majorana neutrino mass matrix \\ with CP symmetry breaking %
\thanks{Presented by B.~Dziewit at the XXXV International Conference of Theoretical Physics "Matter to the Deepest 2011", Ustro\'n, Poland, September 12--18, 2011.}%
}
\author{Bartosz Dziewit, Sebastian Zajac, Marek~Zralek \\
\small{Institute of Physics, University of Silesia, Poland}
}
\date{}
\maketitle
\begin{abstract}
From the new existing data with not vanishing  $\theta_{13}$ mixing angle we determine the possible shape of the  Majorana neutrino mass matrix. We assume that CP symmetry is broken and all Dirac and Majorana phases are taken into account. Two possible approaches "bottom-up" and "top down" are presented. The problem of unphysical phases is examined in detail.
\end{abstract}
%\PACS{14.60.Pq}
\section{Introduction}
 It is commonly believed that    determination of the shape of neutrino mass matrix could shade a light on a mass generation mechanism and give us some information about model lying behind it.
\par 
Many attempts have been made in literature (e.g. \cite{King:2011zj}
 \cite{Dev:2011jc}, \cite{Lashin:2009yd}, \cite{Lam:2011ag}) to
restrict the form of neutrino mass matrices. In general we can divide them into two categories which are called "top-down" - where  the neutrino masses, the mixing angles and the CP violation phases are predicted from a given mass matrix, and  "bottom-up"  method -  where the existing neutrino data determine possible shape of the mass matrix.
\par
In the "top-down" method the neutrino mass matrix, its textures \cite{Xing:2004ik} and symmetries are predicted from some theory beyond the Standard Model (SM).  And the other way in "bottom-up" approach, from neutrino mass matrix we can find all physical neutrino parameters as well as unphysical phases. For three and four neutrino states we can do everything analytically. For a larger number of neutrinos, only numerical method can be used.
We use the base where the charged lepton mass matrix is diagonal so the unitary matrix which diagonalize neutrino mass matrix is also, with an accuracy to non-physical phases, the ordinary Maki, Nakagawa, Sakata, Pontecorvo (MNSP) mixing matrix. 
\par 
In this paper both method are presented. We will  focus on the simpleast, but still realistic, three dimensional case. For the "bottom-up" case we generalize the approach given in paper \cite{Dziewit:2006cg} were only CP conserving problems were considered. From this paper 
among other things we have learned that  textures of  $\mathcal{M}_{\nu}$  
with number of zeros $n \geqslant 3$  do not reproduce experimental data (at $3 \sigma$ C.L.),
there are seven two zero textures which give results 
in agreement with present  data, some of them can produce 
normal, inverse and degenerate mass hierarchies. 
\par
The important goal of present considerations is to check if the CP breaking case is able to change the properties of the mass matrix which still predicts the correct neutrino parameters (Chapter II).
In the "top -down" approach we give exact formulas for the neutrino masses, mixing angles, Dirac and Majorana mixing phases. We show how the unphysical phases depends on the parametrization of the MNSP mixing matrix (Chapter III). 
At the end we give some conclusion (Chapter IV).
\section{Bottom-up method}
For Majorana neutrinos, which we consider, the mass matrix  $\mathcal{M_{\nu}}$ must be symmetric and to have CP symmetry breaking in the lepton sector, must be also complex .  In general   $N$ dimensional symmetric  matrix  can be described by $\frac{N^2+N}{2}$ independent parameters. 
In our case $(N=3)$ we can have $12$ parameters - six modulus and six phases.   
Such a matrix can be diagonalized by unitary transformation: 
\begin{equation}
\label{1}
m_{diag}=U^{T} \cdot \mathcal{M_{\nu}}\cdot U,   
\end{equation}
where unitary matrix $U$ is parametrized by:
\begin{equation}
\label{2}
U= f\cdot U_{MNSP} \cdot P.
\end{equation}
 $U_{MNSP}$ is the standard  MNSP \cite{MNSP}, \cite{pdg} mixing matrix as for Dirac neutrino:
\begin{equation}
\left(
\begin{array}{ccc}
 c_{12} c_{13} & c_{13} s_{12} & e^{-i \delta } s_{13} \\
 -c_{23} s_{12}-c_{12} e^{i \delta } s_{13} s_{23} & c_{12} c_{23}-e^{i \delta } s_{12} s_{13} s_{23} &
   c_{13} s_{23} \\
s_{12} s_{23}-c_{12} c_{23} e^{i \delta } s_{13} & -c_{23} e^{i \delta } s_{12} s_{13}-c_{12} s_{23} &
c_{13} c_{23}
\end{array}
\right),
\end{equation}
where as usually we use abreviation e.g. $s12 = sin(\theta_{12})$ and so on, and:
\begin{equation}
f=\left( \begin{array}{ccc}
e^{\imath \beta_1} & 0 & 0 \\
0 & e^{\imath \beta_2} & 0 \\
0 & 0 & e^{\imath \beta_3}
\end{array} \right), \
P=\left( \begin{array}{ccc}
1 & 0 & 0 \\
0 & e^{\imath \alpha_1/2} & 0 \\
0 & 0 & e^{\imath \alpha_2/2}
\end{array} \right) .
\end{equation}

For $3$ dimensional case we have
nine physical parameters: three masses $(m_1 , m_2 , m_3)$, three mixing angels $(\theta_{12},\theta_{13},\theta_{23})$ and three phases $\delta$ - Dirac phase, and two Majorana phases: $\alpha_1,\alpha_2$.
\\
Matrix $f$ is  composed  by 3 non-physical and unmeasurable phases~$\beta_i,~i=1,2,3$.
\\
Using the reverse relation to (\ref{1})  we can express all elements (separately imaginary  and real parts) of the mass matrix as a function of: 
$$
\mathcal{(M_{\nu})}_{ik}=f_{ik}\left(\theta_{12}, \theta_{13}, \theta_{23}, m_1,m_2,m_3, \delta, \alpha_1,\alpha_2,\beta_1, \beta_2, \beta_3 \right),$$
and found their minimal and maximal values for current \cite{Schwetz:2011qt}
 experimental data. In such way for the given neutrino mass hierarchy we are able to show each possible area as a function of the lightest neutrino mass. 
Such a distribution shows us for example possible texture zeros regions. \\
Only one plot for $(\mathcal{M_{\nu}})_{ee}$  as a function of the ligthest neutrino  mass $m_1$ for normal mass hierarchy, seperately for the absolute value and the phase is presented in Fig 1.\\ For the other elements see our page \cite{strona}. 
This plot reconstructs the  results obtained in neutrinoless double beta experiments \cite{neutrinoless}. It is clear that the absolute value of $|(\mathcal{M_{\nu}})_{ee}| $ is  larger than zero  (i) everywhere, for vanishing Majorana phase,  and (ii)  for $m_{1} >0.02 eV$, independently  of the values of Majorana phases. The modulus and phase   $\varphi_{ee}$ of that element do not depend on unphysical phases $\beta_{i}$.
\section{Top-down method}  
Here we present some simple method of finding unambiguous analytical  relations between oscillation parameters and mass matrix elements.
First for any Majorana neutrino mass matrix $\mathcal{M_{\nu}}$ we diagonalize the hermitian matrix
\begin{equation}
\label{mm}
\mathcal{H}=\mathcal{M_{\nu}}^\dag \mathcal{M_{\nu}} .
\end{equation}
Such a matrix is diagonalized by unitary transformation:
\begin{equation}
\mathcal{W}^\dag \mathcal{H} \mathcal{W}= \left(
\begin{array}{ccc}
 \text{m}_1^2 & 0 & 0 \\
 0 & \text{m}_2^2 & 0 \\
 0 & 0 & \text{m}_3^2
\end{array}
\right),
\end{equation} 
where  the unitary matrix $\mathcal{W}$ is build from the eigenvectors of $\mathcal{H}$:
\begin{equation}
\mathcal{W}=
\left(
\begin{array}{ccc}
  \text{$x_1$} &  \text{$x_2$} &  \text{$x_3$}
   \\
 \text{$y_1$} &  \text{$y_2$} &  \text{$y_3$}
   \\
  \text{$z_1$} &  \text{$z_2$} &  \text{$z_3$}
\end{array}
\right),
\end{equation}
and eigenvalues $m_{i}^{2}, (i=1,2,3)$  are squares of neutrino masses. The normalized eigenvectors are set out with an accuracy of phase.  We can use that freedom in order to find matrix U which diagonalize $\mathcal{M_{\nu}}$ as in Eq. (\ref{1}) with real and positive eigenvalues $m_{i}$.
\begin{equation}
\label{U}
U=
\left(
\begin{array}{ccc}
e^{i \chi_1} & 0 & 0 \\
 0 & e^{i \chi_2} & 0 \\
 0 & 0 & e^{i \chi_3}
\end{array}
\right)
\left(
\begin{array}{ccc}
  \text{$x_1$} &  \text{$x_2$} &  \text{$x_3$}
   \\
 \text{$y_1$} &  \text{$y_2$} &  \text{$y_3$}
   \\
  \text{$z_1$} &  \text{$z_2$} &  \text{$z_3$}
\end{array}
\right).
\end{equation}
The new phases $\chi_{i}, (i=1,2,3)$ also depends on the element of $\mathcal{M_{\nu}}$. 
Now comparing Eq. (\ref{2}) and (\ref{U}) we can find relations:
\begin{equation}
\sin \theta_{13}=\left|x_3\right|, \ \ \ \quad \cos \theta_{13}=\sqrt{1-\left|x_3\right|^2},
\end{equation}
\begin{equation}
\sin \theta_{23}=\frac{\left|y_3\right|}{\sqrt{1-\left|x_3\right|^2}}, \quad \cos \theta_{23}=\frac{|z_3|}{\sqrt{1-\left|x_3\right|^2}},
\end{equation}
\begin{equation}
\sin \theta_{12}=\frac{\left|x_2\right|}{\sqrt{1-\left|x_3\right|^2}}, \quad \cos\theta_{12}=\frac{|x_1|}{\sqrt{1-\left|x_3\right|^2}},
\end{equation}
\begin{equation}
e^{-i \delta}=\frac{|x_1||x_3||z_3|+|z_1|e^{i\left(\tau_1+\omega_3-\tau_3-\omega_1\right)}}{1-|x_3|^2},
\end{equation}
\begin{equation}
\frac{\alpha_1}{2}=\omega_2-\omega_1,
\end{equation}
\begin{equation}
\frac{\alpha_2}{2}=  \omega_3+\delta -\omega_1,
\end{equation}
\begin{equation}
\beta_1=\chi_1+\omega_1,
\end{equation}
\begin{equation}
\beta_2=\chi_2+\eta_3-\omega_2+\omega_1,
\end{equation}
\begin{equation}
\beta_3=\chi_3-\delta -\omega_3+\tau_3+\omega_1.
\end{equation}
where  $\omega_i=Arg(x_i)$, $\eta_i=Arg(y_i)$, $\tau_i=Arg(z_i)$ respectively and $\omega_1,\tau_3=0 \lor \pi$.
\\
Analytical  details for presented equations  are given in the Appendix.
\section{Conclusions and results.}
We have presented two possible approaches of studies neutrino mass matrix. For "bottom-up" method we have get possible values of $\mathcal{M_{\nu}}$ matrix elements from current experimental data.  As a example we have presented $\mathcal({M}_{\nu})_{ee}$ element which agrees with the neutrinoless double beta decay observations.\\
We have learned that Majorana phases are crucial to get some texture for neutrino mass matrix e.g. zero texture. Any other symmetry imposed on $\mathcal{M_{\nu}}$ can be studied in the same way. Whole set of plots and computer program used for calculations is possible to see on web-page \cite{strona}.
From "top-down" method we have learned how to find all physical neutrino parameters from given neutrino mass matrix which follow from any physics beyond the SM. This knowledge is useful in future plan context. We would like to enlarge our analytical solutions for $3+1$ mass matrix case and numerical solutions for $6 \times 6$  dimensional $\mathcal{M_{\nu}}$ (i.e like  presented  in \cite{Xing:2011ur}).
\begin{figure}[!hbt]
	\centering
	\includegraphics[scale=0.37]{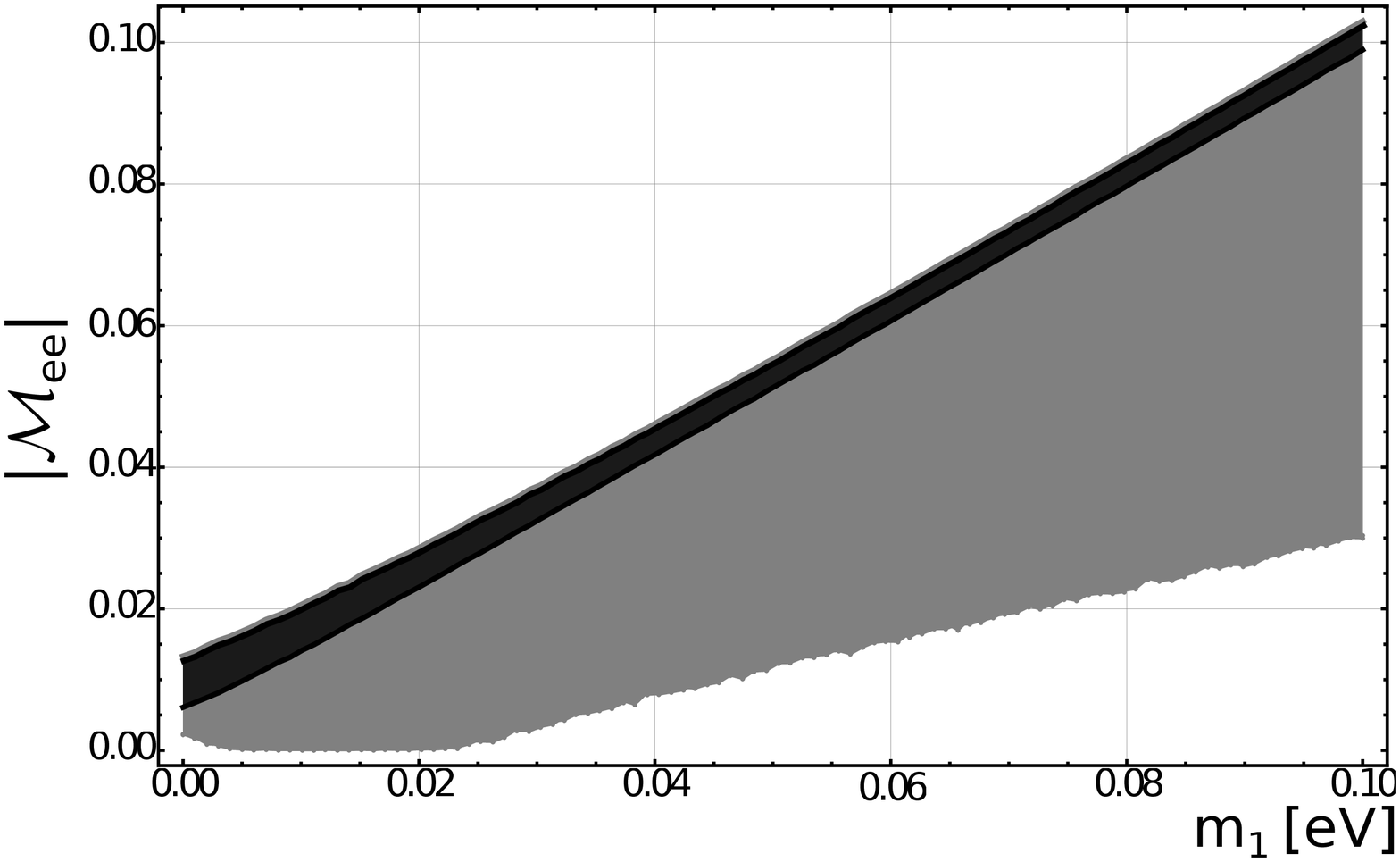}
	\centering
	\includegraphics[scale=0.37]{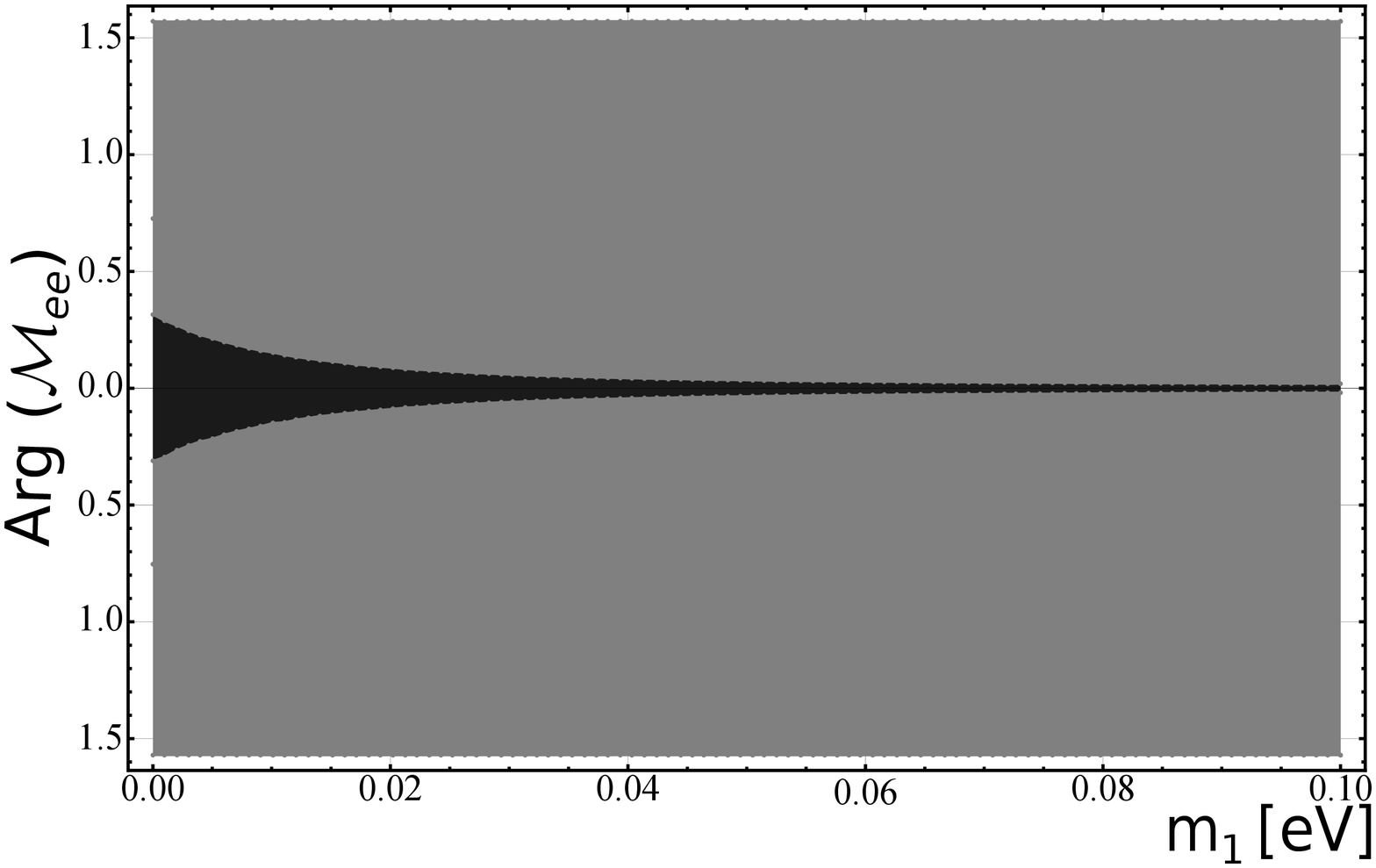}
\caption{Allowed values of modulus and phase  (upper and lower figure respectively) of $\mathcal{M}_{ee}$ for normal mass hierarchy as a function of lightest neutrino mass $m_1$. Darker region shows  part with $\alpha_1,\alpha_2=0$ and $\delta\neq 0$. Lighter one shows part with $\alpha_1,\alpha_2, \delta\neq 0$. Plot was made for $10^{6}$ randomly generated oscillation parameters at $2 \sigma$ C.L.}
\end{figure}
  \newpage
\section*{Appendix}
 We would like to give here formulas which express the physical neutrino parameters by elements of the mass matrix. Let us parametrize the $\mathcal{H}$ matrix in the way:
 \begin{equation}
 \mathcal{H}=
\left(
\begin{array}{ccc}
 A & B e^{i \phi_1 }& C e^{i \phi_2} \\
 B e^{-i \phi_1} & D & E e^{i \phi_3} \\
 C e^{-i \phi_2} & E e^{-i \phi_3} & F
\end{array}
\right),
\end{equation}
From Eq. (\ref{mm}) each element of the $\mathcal{H}$ can be easy expressed by the modulus and phases of $(\mathcal{M}_{\nu})_{a,b}=m_{a,b}  \  e^ { i \varphi_{a,b}}, (a,b=e,\mu,\tau) $.
\ 
The matrix $\mathcal{H}$ eigenvalues are given by:
\begin{equation}
m_1^2=\frac{2}{3} p \cos (\phi )-\frac{a}{3},
\end{equation}
\begin{equation}
m_2^2=-\frac{a}{3}-\frac{1}{3} p \left(\cos (\phi )-\sqrt{3} \sin (\phi
   )\right),
\end{equation}
\begin{equation}
m_3^2=-\frac{a}{3}-\frac{1}{3} p \left(\cos (\phi )-\sqrt{3} \sin (\phi
   )\right),
\end{equation}
where:
\begin{equation}
p=\sqrt{a^2-3 b}, \quad \phi=\frac{1}{3} \arccos \left(-\frac{1}{p^2} \left( a^3  - \frac{9}{2}ab +\frac{27}{2}c  \right)\right),
\end{equation}
and:
\begin{equation}
a=-Tr [\mathcal{H}],
\end{equation}
\begin{equation}
b=AD+AF+DF-B^2-C^2-E^2,
\end{equation}
\begin{equation}
c=AE^2+DC^2+FB^2-ADF-2BCE \cos \left(  \phi_1 +\phi_3 - \phi_2 \right).
\end{equation}
\
The normalized $\mathcal{H}$ eigenvectors are given by:
\begin{equation}
\left(
\begin{array}{c}
 x_1 \\
 y_1 \\
 z_1
\end{array}
\right)=
\frac{1}{\sqrt{X_1^2+\left|Y_1\right|{}^2+\left|Z_1\right|{}^2}} \cdot
   \left(
\begin{array}{c}
 X_1 \\
 Y_1 \\
 Z_1
\end{array}
\right),
\end{equation}
\begin{equation}
\begin{array}{c}
 X_1=(D-m_1^2) (F-m_1^2)-E^2, \\
 Y_1=C\cdot E e^{-i (\phi_2 -\phi_3 )}-B e^{-i \phi_1 } (F-m_1^2), \\
 Z_1=B\cdot E e^{-i (\phi_1 +\phi_3 )}-C e^{-i \phi_2 } (D-m_1^2),
\end{array}
\end{equation}
\begin{equation}
\left(
\begin{array}{c}
 x_2 \\
 y_2 \\
 z_2
\end{array}
\right)=
\frac{1}{\sqrt{\left|X_2\right|{}^2+Y_2^2+\left|Z_2\right|{}^2}} \cdot
   \left(
\begin{array}{c}
 X_2 \\
 Y_2 \\
 Z_2
\end{array}
\right),
\end{equation}
\begin{equation}
\begin{array}{c}
 X_2=C \cdot E e^{i (\phi_2 -\phi_3 )}-B e^{i \phi_1 } (F-m_2^2), \\
 Y_2=(A-m_2^2) (F-m_2^2)-C^2, \\
 Z_2=B\cdot C e^{i (\phi_1 -\phi_2 )}-E e^{-i \phi_3 } (A-m_2^2),
\end{array}
\end{equation}
\begin{equation}
\left(
\begin{array}{c}
 x_3 \\
 y_3 \\
 z_3
\end{array}
\right)=
\frac{1}{\sqrt{\left|X_3\right|{}^2+\left|Y_3\right|{}^2+Z_3^2}} \cdot
   \left(
\begin{array}{c}
 X_3 \\
 Y_3 \\
 Z_3
\end{array}
\right),
\end{equation}
\begin{equation}
\begin{array}{c}
X_3= B\cdot E e^{i (\phi_1 +\phi_3 )}-C e^{i \phi_2 } (D-m_3^2), \\
Y_3= B  \cdot C e^{-i (\phi_1 -\phi_2 )}-E e^{i \phi_3 } (A-m_3^2), \\
Z_3=(A-m_3^2) (D-m_3^2)-B^2.
\end{array}
\end{equation}
 
\end{document}